\documentclass[twoside]{optica-article}
\journal{opticajournal} 
\articletype{Research Article}

\usepackage{lineno}
\usepackage[utf8]{inputenc}
\usepackage{array}
\usepackage{makecell}
\usepackage{multirow}
\usepackage{caption}
\usepackage[subrefformat=parens,labelformat=parens]{subcaption}
\usepackage{fancyhdr}

\usepackage{comment}

\usepackage[capitalize]{cleveref}

\crefrangelabelformat{subfigure}%
{ #3\crefstripprefix{#2}{#1}#4--#5\crefstripprefix{#1}{#2}#6)}

\begin{document}

\title{Ultra-narrow linewidth laser across the C-band using polarization-controlled dual-cavity feedback}
\author{Jeppe H. Surrow,\authormark{1,$\dagger$,*} 
Simon T. Thomsen,\authormark{1,$\dagger$} 
Rakesh R. Kumar,\authormark{1} 
Mónica Far Brusatori,\authormark{1} 
Maria Paula Montes,\authormark{1} 
Ahan S. Palsole,\authormark{1}
Chris Hoede,\authormark{2} 
Holger N. Klein,\authormark{3} 
and Nicolas Volet\authormark{1}}

\address{\authormark{1}Department of Electrical and Computer Engineering, Aarhus University, 8200 Aarhus N., Denmark\\
\authormark{2}AimValley, 1213TV Hilversum, the Netherlands\\
\authormark{3}OE Solutions America, Goleta, CA 93117, USA\\}
\affil{\authormark{$\dagger$}These authors contributed equally to this work}
\email{\authormark{*}surrow@ece.au.dk} 

\begin{abstract}
    A standard method to reduce the linewidth of semiconductor lasers involves the use of external optical feedback (EOF). However, feedback powers less than 1~\% usually trigger coherence collapse (CC), leading to chaotic laser dynamics and linewidth broadening. This paper explores a method to mitigate CC through precise tuning of the feedback polarization depending on the feedback power. We report a semiconductor laser with a sub-100~Hz intrinsic linewidth, achieved via EOF. The laser features a U-shaped cavity with two sampled grating distributed Bragg reflectors (SG-DBRs), enabling broad tunability across a 42 nm wavelength range (1513--1555 nm). By injecting optical feedback into both sides of the laser cavity via an external fiber-based cavity, we reduce the intrinsic linewidth by more than three orders of magnitude, from MHz to sub-kHz across the laser's tuning range. By dynamically tuning the polarization, we demonstrate sub-100~Hz intrinsic linewidths at feedback powers up to 10~\%, marking an improvement over prior studies where CC limited performance.
\end{abstract}
\section{Introduction}

Narrow-linewidth semiconductor lasers with broad wavelength tunability are critical components in a range of advanced technologies, including optical communications \cite{jorgensen2022petabit,seimetz2008laser}, 
quantum photonics \cite{holewa2024high,Avesani_2021_Integrated_QKD}, 
LiDAR \cite{riemensberger2020massively},
and spectroscopy \cite{yang2019vernier}, 
all of which benefit from chip-scale integration. 
In coherent optical communications,  
advanced modulation formats such as 128 quadrature amplitude modulation (QAM) require lasers with linewidths narrower than 100 kHz to support increasing data transmission rates \cite{jones2018stop,hamaoka2019ultra}. Although such linewidths are readily provided on-chip by Brillouin lasers\cite{Gundavarapu_Brillouin_2019} or ring-lasers\cite{far2022ultralow}, these lasers are limited in tunability across the C-band that is essential for dense wavelength division multiplexing (DWDM), a key enabler in modern optical networks.
For DWDM, an array of non-tunable or narrow bandwidth tunable chip lasers\cite{nie2024simultaneous} could be sufficient, but the precise manufacturing required to obtain the specific wavelengths for each channel complicates this.
A simpler approach is to have a single tunable laser, such as an SG-DBR laser \cite{zhao2018indium}, which is much easier to replace and is already mature in the integrated tunable laser assembly market. 
However, its intrinsic linewidth, typically ranging from hundreds of kHz to several MHz \cite{renaudier2022devices}, still requires further reduction.

The effective linewidth and the reduction thereof is also important for applications requiring long-term stability such as coherent communication schemes \cite{Fang_Overcoming_2024}. Having low technical noise in the laser is therefore important as well.
\thispagestyle{fancy}

\fancyfoot{} 
\fancyfoot[RO]{\hypersetup{linkcolor=black}
\fontsize{8pt}{9pt}\sffamily{\hypersetup{urlcolor=black}\url{https://doi.org/10.1364/OE.544372}}\\\fontsize{8pt}{9pt}\sffamily Received 8 Oct 2024; revised 3 Feb 2025; accepted 11 Feb 2025; published 6 Mar 2025}
\fancyfoot[LO]{\fontsize{8pt}{9pt}\sffamily\#544372\\\fontsize{8pt}{9pt}\sffamily Journal \copyright\, \number\year}
Several techniques have been explored for narrowing intrinsic and effective laser linewidth. Electronic stabilization methods such as Pound-Drever-Hall (PDH) locking \cite{guo2022chip} and feed-forward techniques \cite{idjadi2020nanophotonic} reduces the effective laser linewidth.
While the aforementioned methods succeed in reducing the effective linewidth, they often require complex setups with reference cavities or phase discriminators, adding to system cost and complexity.
Reduction of both intrinsic and effective linewidth can be achieved through optical feedback methods\cite{guo2022chip}, which offer a simpler and more cost-effective solution by re-injecting part of the laser’s output back into the cavity.

Such feedback can be provided by frequency-selective elements like Fabry-Perot cavities \cite{lewoczko2015ultra} and
diffraction gratings \cite{shin2016widely}, 
or through on-chip components like DBR gratings \cite{kumar202210} 
and micro-ring resonators for self-injection locking \cite{Tran_2022_Extending,prokoshin2024ultra,Isichenko_Sub_2024}.
However, the fabrication of high-Q cavities can be costly and requires careful alignment and frequency-selective components limit the tunability of the laser.

A simpler approach to provide optical feedback is through fiber-based delays 
\cite{Huang_Dual_2017,Brunner_Six_orders_2017}. Compared to chip-based delays, optical fibers can provide feedback delays of several meters with negligible propagation loss, yielding long external cavity lifetimes that are crucial for linewidth narrowing.
For small amounts of coherent optical feedback, i.e feedback polarization aligned with the lasing mode, increasing feedback power tends to narrow the linewidth until the onset of coherence collapse (CC), where chaos sets in and the linewidth broadens
\cite{petermann1995external}.
In quantum-well semiconductor lasers, research has shown that CC typically sets in at feedback powers less than 1~\% of the laser's output \cite{donati2012diagram}, 
making them highly susceptible to small amounts of EOF. Reflections caused by coupling to optical fibers and connectors can readily provide such feedback levels that are detrimental to laser performance.
This necessitates the use of bulky optical isolators in the packaging of the laser diodes to mitigate CC. Incoherent feedback, for which the feedback polarization is orthogonal to the lasing field, has been extensively studied for applications like pulse generation and random bit generation \cite{cheng2003generation, oliver2011dynamics} but also for linewidth reduction via negative optical feedback\cite{yasaka1991fm}.
Although incoherent feedback still indirectly affects the lasing field via the carrier density, it is more robust against high feedback levels \cite{ju2005dynamic}.

In this work, 
we present a significant advancement by combining dual-cavity optical feedback with polarization control to achieve sub-kHz linewidths across the C-band and part of the S-band.
This approach circumvents CC by tuning the feedback polarization relative to the laser output, enabling feedback levels up to 10~\% while maintaining sub-100~Hz linewidths.
The approach extends the region of feedback levels for the minimum-achievable linewidth, otherwise only obtainable in a narrow region on the verge of chaos. 
The laser, based on a monolithically integrated InP photonic circuit, features a U-shaped cavity with two SG-DBR mirrors \cite{Brusatori_Dual_wavelength_2023}, 
providing tunability across a 42 nm wavelength range (1513--1555 nm),
and 10 mW fiber-coupled output power, with a 
side-mode suppression ratio (SMSR) exceeding 35~dB. 
By injecting optical feedback into both sides of the laser cavity through an external fiber-based cavity, we reduce the intrinsic linewidth by more than four orders of magnitude, achieving intrinsic linewidths as low as 13 Hz. The effective linewidth is reduced by two orders of magnitude, from several MHz to tens of kHz.
This marks a significant improvement over previous U-shaped laser studies, which were constrained by CC. To the best of our knowledge, this is the first study to achieve sub-100~Hz linewidths at feedback levels beyond CC through adaptive polarization rotation. Our results are consistent with or better than recent studies on optical feedback in SG-DBR lasers,
demonstrating that complex filtering techniques like microrings
\cite{Sheng_2024_Advances}
or interference phenomena \cite{wang2024widely}
are not necessarily needed.
Lastly, we conduct an analysis of the relaxation oscillation frequency ($f_\mathrm{RO}$) based on relative intensity noise (RIN) spectra for different laser gains, feedback levels, and polarizations. Our findings show robustness against CC for higher laser gains and confirm the effectiveness of misaligning the polarization to avoid CC.

\section{Theoretical background}\label{section: Theoretical background}
Laser linewidth reduction using EOF is a well-known technique that serves as the basis for this study. Coupling to an external cavity increases the lifetime of coherent photons in the lasing field, causing a reduction in the intrinsic linewidth. For a given ratio between the laser output power $P_0$ and the optical feedback power $P_{\mathrm{ext}}$,  defined as $f_{\mathrm{ext}} = P_{\mathrm{ext}}/P_0$, the intrinsic linewidth can be modeled as \cite{petermann1995external}
\
\begin{equation}
    \Delta f = \frac{\Delta f_0}{\left[1 + \frac{\tau_e}{\tau_L} \beta  \sqrt{f_{\mathrm{ext}}} \cos{\left(\phi_e + \arctan{\alpha} \right)} \right]^2},
    \label{eq:feedback}
\end{equation}
\
where $\Delta f_0$ is the intrinsic linewidth of the solitary laser ($P_{\mathrm{ext}} = 0$), $\tau_L$ is the laser cavity roundtrip time, $\tau_e$ is the external feedback round trip time, and $\alpha$ is the linewidth enhancement factor \cite{henry1982theory}. The parameter $\beta$ describes the coupling strength between the EOF and the laser cavity which depends on the mirror reflectivities and on the polarization overlap between the EOF and the laser output \cite{yasaka1991fm} (i.e. the fraction of coherent feedback). Keeping the external feedback phase $\phi_e = \omega \tau_e$ constant, where $\omega$ is the lasing frequency of the solitary laser, \cref{eq:feedback} suggests that a reduction in the laser linewidth can be achieved by increasing the fraction of optical feedback $f_{\mathrm{ext}}$. 
This has a limit, however, as eventually, this leads to the onset of CC \cite{lenstra1985coherence}, where the linewidth broadens significantly. 
Incoherent feedback, although not contributing to linewidth reduction according to \cref{eq:feedback} as it is orthogonal to the output field ($\beta = 0$), also influences the laser dynamics. However, the level of incoherent feedback that can be sent into the laser cavity before going into a chaotic regime is significantly higher than for coherent feedback \cite{ju2005dynamic}.

\Cref{eq:feedback} is derived assuming feedback from only one side of the cavity \cite{lang1980external}. The laser studied here has feedback from both sides of the laser cavity, where the dynamics can be modeled by the Lang-Kobayashi equation with an additional feedback term\cite{de2024clarifying}. In this case, the feedback is the superposition of the two feedback terms from either cavity, which is analyzed in-depth in Ref. \cite{far2022dynamics}. In this work, we have studied the particular case of dual feedback with equal delays and constructive interference, that can effectively be modeled as a single feedback term, which allows the use of \cref{eq:feedback}.

The main goal of this work is to study how different levels of feedback and polarization impact laser linewidth. We use several methods to measure the intrinsic laser linewidth. One method is the self-coherent envelope linewidth detection (SCELD) method\cite{Mercer_SCELD_1991,Zhao_2022_Narrow}, which is a special case of delayed self-heterodyne (DSH) detection. Contrary to the conventional DSH method \cite{Okoshi_Novel_1980}, this variant employs delays shorter than the laser coherence length. The SCELD method 
offers better accuracy than conventional DSH 
techniques, particularly due to mitigating the effects of \(1/f\) noise.
The DSH spectrum of a purely white noise laser is described in Ref. \cite{Zhao_2022_Narrow} as
\begin{equation}
    S(f) = \frac{P^2 \Delta f}{4 \pi} 
    \frac{1 - e^{-2 \pi \tau \Delta f} \left[\cos(2 \pi f \tau) + \frac{\Delta f}{f} \sin(2 \pi f \tau) \right]}{(\Delta f)^2 + f^2} ,
    \label{eq: zeta function}
\end{equation}
where $P$ is the signal power, $\Delta f$ is the intrinsic linewidth, $f$ is the frequency offset from the modulation frequency $f_s = 76$ MHz, and $\tau=nL/c$ is the time delay, with $L$ being the delay length, $n$ the fiber group index, and $c$ the speed of light in vacuum. For the SCELD method, where $\tau < (2 \pi \Delta f)^{-1}$, the exponential term becomes significant and sinusoidal ripples are present in the spectrum. The intrinsic linewidth is determined by fitting \cref{eq: zeta function} to these ripples, the details of which are explained in the following section. The second method is the coherent self-heterodyne (CSH) method \cite{Thomsen_CSH_2023}, which typically employs even shorter delays than the SCELD method. Using delays much shorter than the laser coherence length, the signal, captured on an electrical spectrum analyzer (ESA) is directly proportional to the lasers frequency noise (FN) power spectral density (PSD). We obtain the intrinsic linewidth directly by multiplying the FN PSD noise floor by $\pi$. The last way to measure the intrinsic linewidth is by using a commercial frequency noise analyzer (FNA, HighFinesse, LWA-1k-1550). This instrument also measures the FN PSD.
\section{Experimental setup and methods} 
The experimental setup features a U-shaped laser which enables dual cavity feedback via a single EOF loop, and a measurement branch for precise linewidth characterization. With this setup, we aim to study the linewidth-narrowing effect of EOF across the lasers wavelength tunability range.
The following sections explain in depth the different components of the measurement methods.

\subsection{Laser configuration}
\begin{figure}[t]
    \centering
    \begin{subfigure}[b]{0.45\textwidth}
        \caption{}
        \centering
        \includegraphics[width = \textwidth]{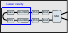}
        \label{fig:U-shaped laser}
    \end{subfigure}
    \hfill
    \setcounter{subfigure}{2}
    \begin{subfigure}[b]{0.5\textwidth}
        \caption{}
        \centering
        \includegraphics[width = \textwidth]{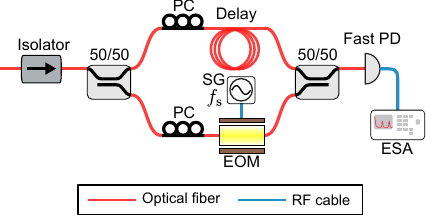}
        \label{fig: Meas setup 2}
    \end{subfigure}
            \hfill
    \setcounter{subfigure}{1}
    \begin{subfigure}[b]{0.7\textwidth}
        \vspace{-10pt}
        \caption{}
        \centering
        \includegraphics[width = \textwidth]{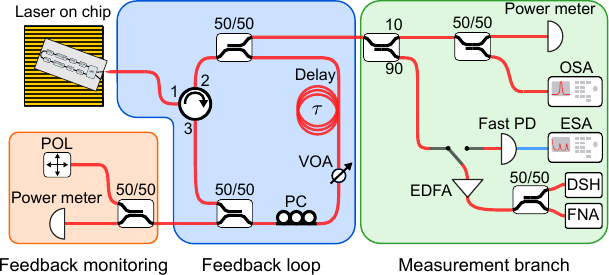}
        \label{fig: Meas setup 1}
    \end{subfigure}
    \caption{Schematic of (a) the U-shaped laser with the laser cavity highlighted by a blue dashed line box, SOA: semiconductor optical amplifier, SG-DBR: sampled grating - distributed Bragg reflector, PhS: phase shifter, MMI: multi-mode interferometer, (b) the measurement setup consisting of a feedback loop (blue box), feedback monitoring (orange box), and a measurement branch (green box), PC: polarization controller, POL: polarimeter, VOA: variable optical attenuator, EDFA: erbium-doped fiber amplifier, DSH: delayed self-heterodyne method, OSA: optical spectrum analyzer, and (c) the DSH measurement method, EOM: electro-optic modulator, PD: photodetector, ESA: electrical spectrum analyzer, FNA: Frequency noise analyzer.}
    \label{fig:Setup}
\end{figure}

The U-shaped laser\cite{klein2022method,blumenthal2017tunable} used in this work is a monolithically integrated InP photonic integrated circuit (PIC), which is shown schematically in \cref{fig:U-shaped laser}.
The laser cavity (blue dashed line box) consists of a gain section and a phase shifter situated between two SG-DBR mirrors allowing tunability across the C-band and partially across the S-band. Light is emitted from both mirrors and combined by a 2$\times$2 multi-mode interferometer (MMI) into the output waveguides of the PIC. Phase shifters are included in both arms to ensure constructive interference at the MMI, maximizing the output power. This simultaneously ensures constructive interference for the dual cavity feedback. Semiconductor optical amplifiers (SOAs) are also present in the output arms for further power amplification.
The laser is mounted on a heat-sinking carrier connected to a custom current controller for tunability and optimization. A Peltier element beneath the carrier stabilizes the temperature at 23.500 $\pm$ 0.003 \textdegree C, yielding a wavelength tunability range of 1513--1555~nm.

\subsection{Optical feedback setup}

The setup used to inject optical feedback into the laser is shown in \cref{fig: Meas setup 1}. The output laser light of the PIC is collected by an anti-reflection coated lensed fiber which is connected to a circulator. A $50/50$ directional coupler then splits the light into two branches. One branch is the feedback loop (blue box) and the other branch is the measurement branch (green box). In the former, a variable optical attenuator (VOA, Thorlabs VOA50-APC) is used to control and adjust the optical feedback strength. A polarization controller (PC) is used to control the polarization of the feedback. After passing through these components, the light is then split $50/50$ again. One output provides optical feedback to the laser. The other output (orange box) is split 50:50 into a sensitive power meter (Thorlabs S154C) and a polarimeter (Thorlabs PAX1000IR2) to monitor the feedback power and polarization, respectively. In the measurement branch, 10~\% is tapped off to measure the laser output power and wavelength by a power meter (Thorlabs S132C) and optical spectrum analyzer (OSA, Yenista OSA20), respectively. The remaining 90~\% is either coupled into a fast photodetector (HP 11982A, 15~GHz bandwidth) going to an electrical spectrum analyzer (ESA, R\&S FSW50) for RIN measurements or passes through an erbium-doped fiber amplifier (EDFA, Lumibird-Keopsys CEFA) to ensure sufficient power for the DSH configuration (schematized in \cref{fig: Meas setup 2}) and the frequency noise analyzer (FNA, HighFinesse LWA-1k-1550).

\subsection{Self-coherent envelope linewidth detection and frequency noise spectra}\label{subsection: DSH measurement method}
The DSH setup used for the SCELD method is shown in \cref{fig: Meas setup 2}, where the input light is split $50/50$, sent through PCs to maximize the output of the electro-optic modulator (EOM, iXblue MPZ-LN-01) and subsequently optimize the constructive interference when recombining the unbalanced arms. This is then measured by a fast photodetector (Thorlabs PDA05CF2) and analyzed by an ESA (R\&S FSW50).

\begin{figure}[hptb]
    \centering
    \begin{subfigure}[b]{0.5\textwidth}
        \caption{}
        \centering
        \includegraphics[width = \textwidth]{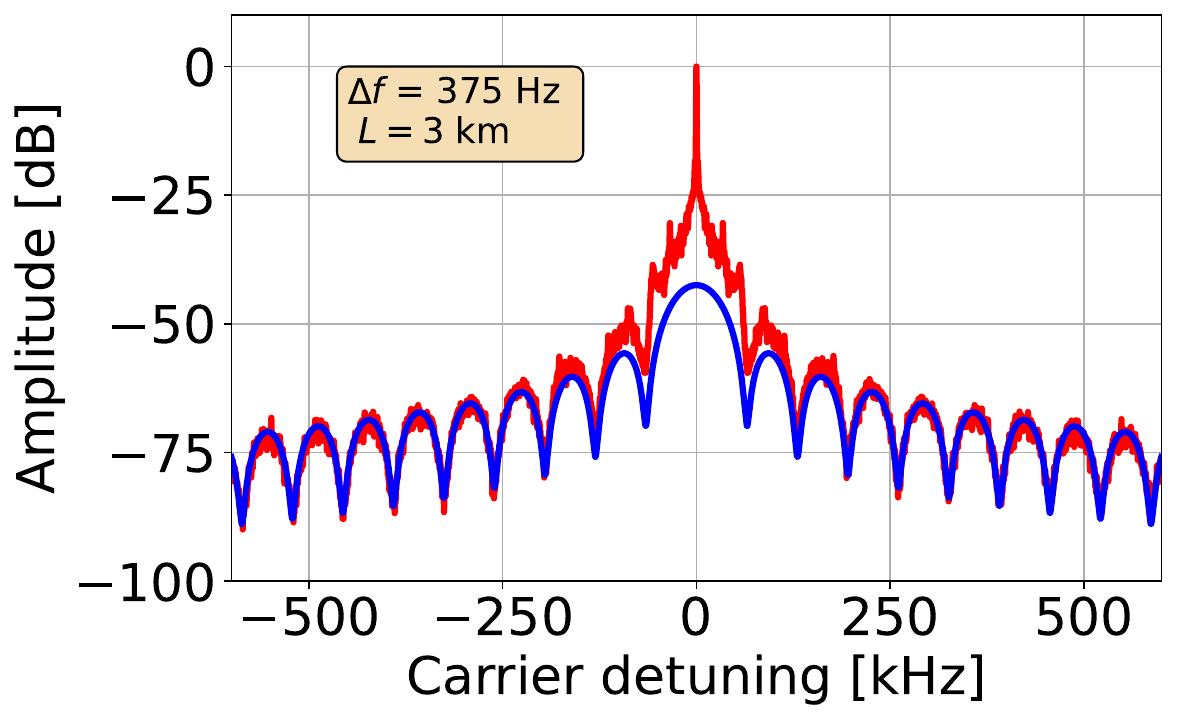}
        \label{fig:zeta_fit_3000}
    \end{subfigure}
    \hfill
    \begin{subfigure}{0.48\textwidth}
        \caption{}
        \centering
        \includegraphics[width=\linewidth]{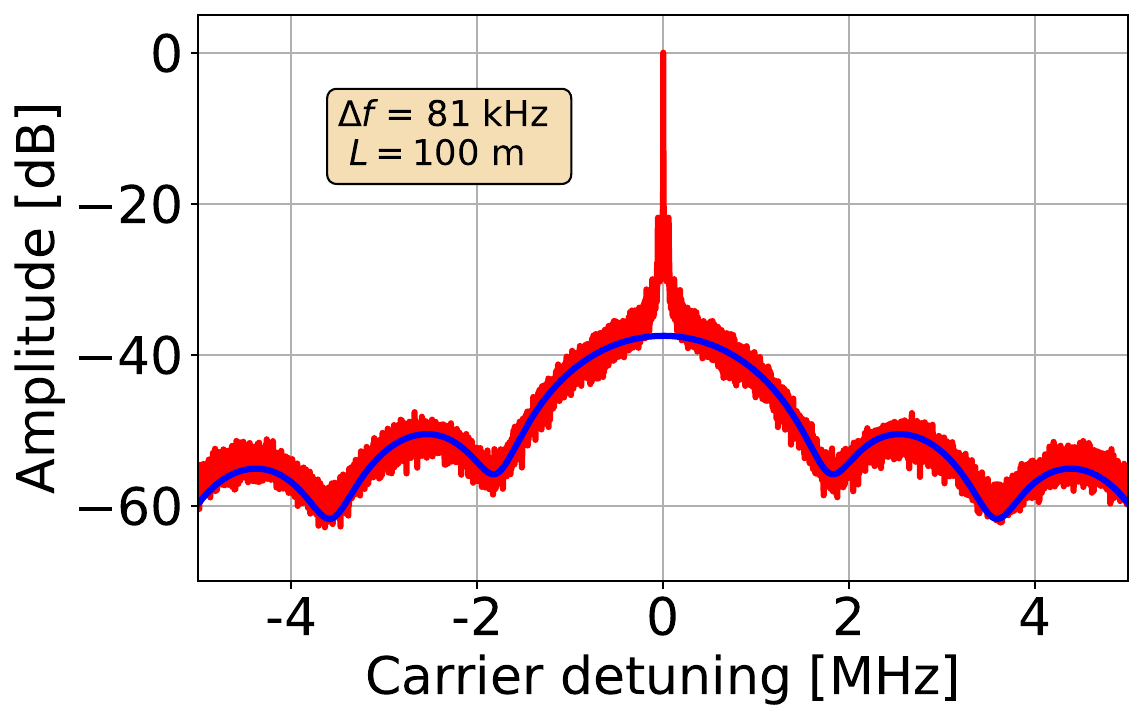}
        \label{fig:zeta_fit_100}
    \end{subfigure}


    
    \begin{subfigure}[b]{0.48\textwidth}
        \vspace{-20pt}
        \caption{}
        \centering
        \includegraphics[width = \textwidth, height=0.68445\textwidth]{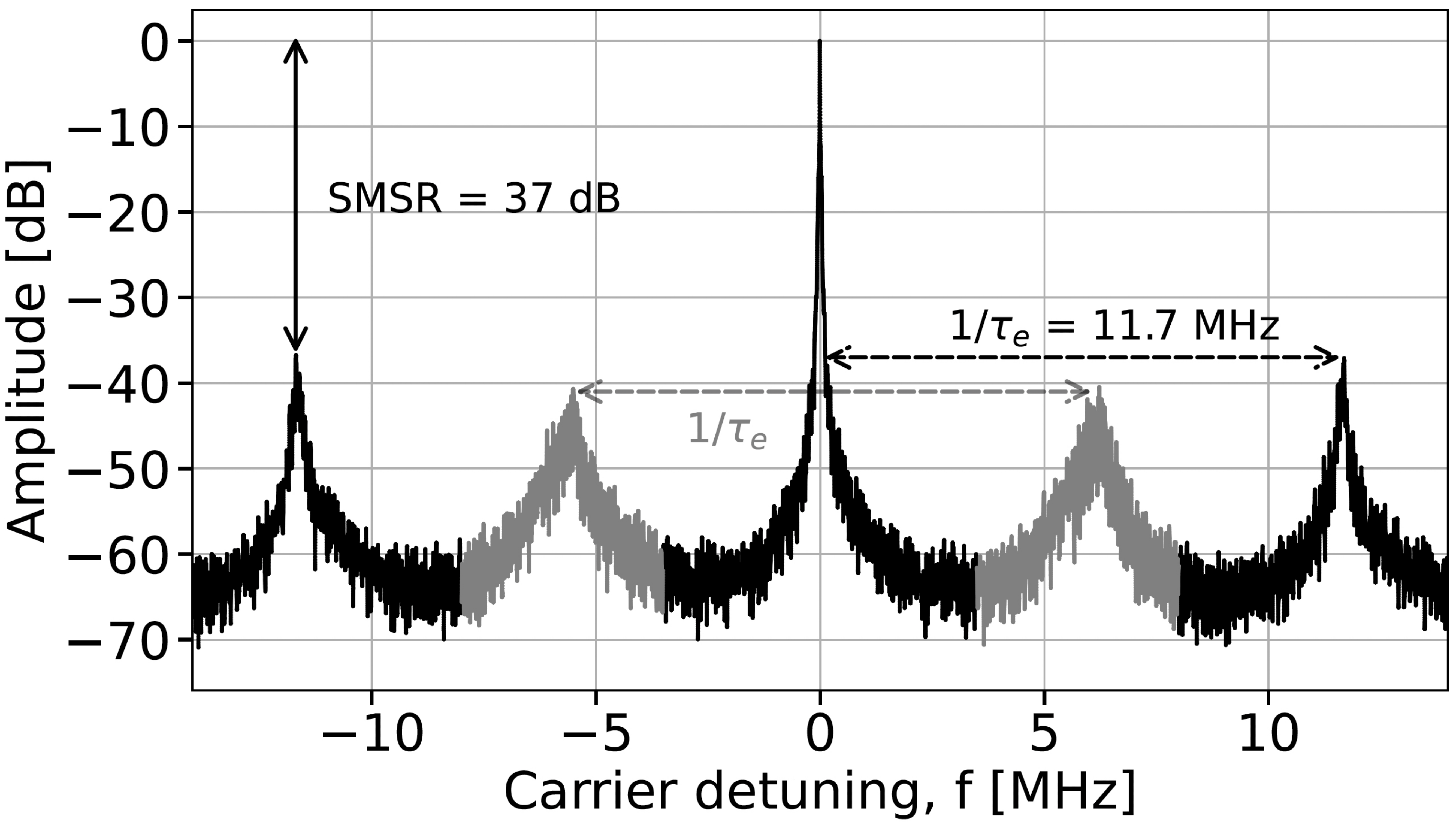}
        \label{fig: SMSR spectrum}
    \end{subfigure}
    \hfill
    \begin{subfigure}[b]{0.48\textwidth}
        \vspace{-20pt}
        \caption{}
        \centering
        \includegraphics[width = \textwidth,height=0.68445\textwidth]{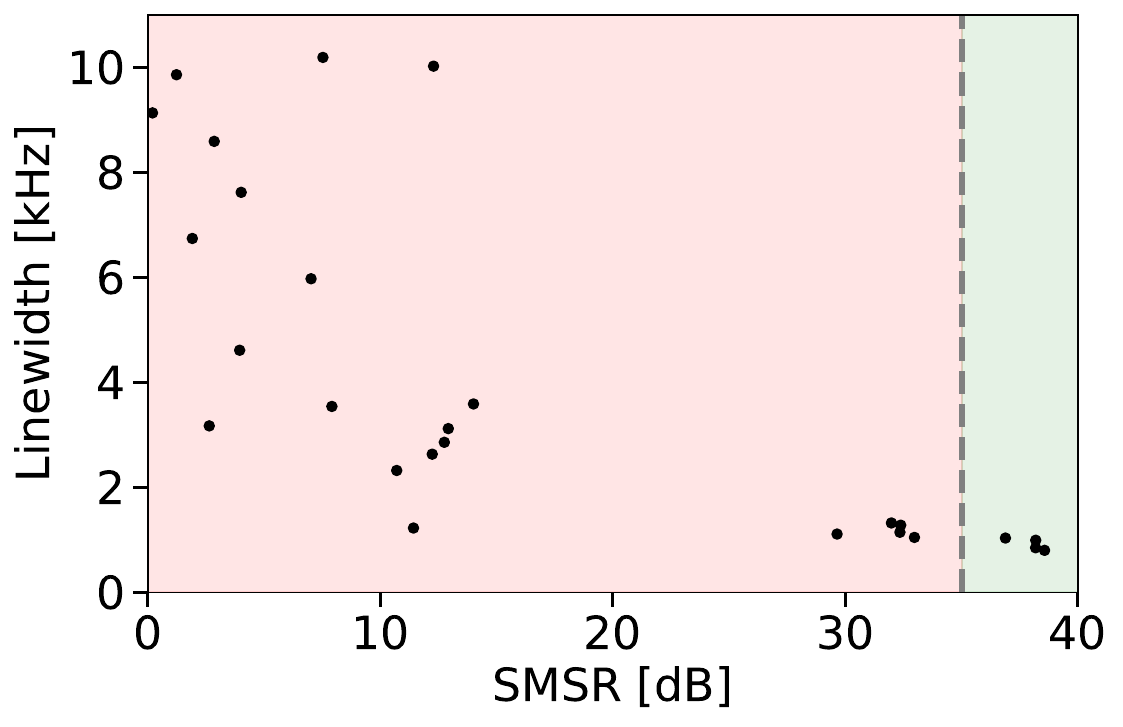}
        \label{fig: SMSR vs linewidth}
    \end{subfigure}
    \caption{DSH traces fitted using the SCELD method with delays of (a) 3~km and (b) 100~m.
    (c) ESA spectrum highlighting the external cavity mode spacing of $1/\tau_e = 11.7$~MHz and an SMSR of $37$~dB. Highlighted in grey are the external cavity modes from the DC part of the spectrum. (d) SMSR against linewidth for a $-29$~dB feedback ratio. The single-mode regime is determined to be for SMSR $>$ $35$~dB.}
    \label{fig:Zeta_fits}
\end{figure}

Examples of \cref{eq: zeta function} being used to extract linewidths, ranging from MHz to sub-kHz, with different delays are shown
in \cref{fig:zeta_fit_3000,fig:zeta_fit_100}. The periodic ripples are separated by the inverse of the interferometer delay $1/\tau \propto 1/L$. Close to the carrier frequency (<200 kHz), the spectra are not well modeled by \cref{eq: zeta function}, where $1/f$ noise, other features from the FN power spectral density (PSD), and a delta peak appear \cite{Thomsen_CSH_2023}. This is especially evident in \cref{fig:zeta_fit_3000}. Consequently, the spectra are fitted only for carrier detunings above $200$ kHz. For a given delay $L$, the linewidth is determined by the height of these ripples. Deeper ripples indicate a more coherent signal and thus a lower linewidth as mentioned in \cref{section: Theoretical background}.

For a given linewidth, the delay should be chosen accordingly, as the amplitudes of the ripples depend on both. If the delay is too long the ripples disappear. On the other hand, a delay that is too short yields ripples that extend below the measurement noise floor. For example, a $L = 3$ km delay can be used for linewidths between approximately $100$ Hz to $10$ kHz, whereas a $L = 30$ m delay can measure linewidths ranging from $200$~kHz to several MHz.

The CSH method that is used to obtain the FN spectrum uses the same setup as the SCELD method but with a 30 m delay to stay within the coherent regime at all times. ESA spectra are averaged 50 times using a 100 Hz resolution bandwidth and 5 MHz span. Lastly, we obtain the spectra from the FNA that are averaged 10 times with a 50 Hz resolution bandwidth and a 10 MHz span. 

\subsection{Stability}\label{subsection: Stability}
Linewidth narrowing using EOF depends critically on the overlap between the solitary cavity lasing mode and the external cavity modes (ECMs) due to the optical feedback path \cite{petermann1995external}. 
The ECMs appear on the ESA spectrum as a beat note between the lasing mode and neighboring ECMs, as exemplified in \cref{fig: SMSR spectrum}.
The $18$~m fiber feedback loop gives rise to a relatively narrow ECM separation with a free spectral range of $1/\tau_e = 11.7$ MHz. By measuring the beat-note between the U-shaped laser with no optical feedback and another laser (Pure Photonics PPCL550), we observe a frequency drift corresponding to approximately 1 GHz over the course of an hour. This drift is attributed to ambient temperature fluctuations in the laboratory environment, which affect both the fiber feedback loop and the laser itself. This leads to misaligned cavity modes and eventually mode-hops to a neighboring ECM. In terms of \cref{eq:feedback} this mode overlap is modeled by the external feedback phase $\phi_e$, from which a maximum linewidth reduction is achieved when the external feedback interferes constructively with the laser cavity field. In order to investigate how the overlap between the lasing mode and ECMs affects the linewidth, we monitored the SMSR and intrinsic linewidth from the SCELD method continuously for one minute as shown in \mbox{\cref{fig: SMSR vs linewidth}}. We observe a significant linewidth broadening when the SMSR is low. There is also a gap between $15$~dB and $28$~dB SMSR. This gap can be explained by the onset of mode competition as the solitary laser cavity mode drifts in between two ECMs and there is a sudden jump in SMSR. Based on this, all our subsequent measurements using the SCELD method will be done for SMSR $>$ $35$~dB to make sure that the laser remains single mode during linewidth measurements. Additionally, this ensures constructive feedback between the feedback and laser cavity fields, effectively eliminating the $\phi_e$ dependence in \cref{eq:feedback}. This allows us to investigate how the linewidth evolves with feedback power without needing to take $\phi_e$ drift into account.

\section{Results and Discussion}

In this section, we present the experimental results and analysis of linewidth reduction and laser dynamics achieved through polarization-controlled EOF. Our focus lies in examining how feedback power influences the laser's linewidth, relaxation oscillations, and overall output stability.
\begin{figure}[hptb]
    \centering
    \begin{subfigure}[t]{\textwidth}
        \centering
        \caption{}
        \includegraphics[width = \textwidth]{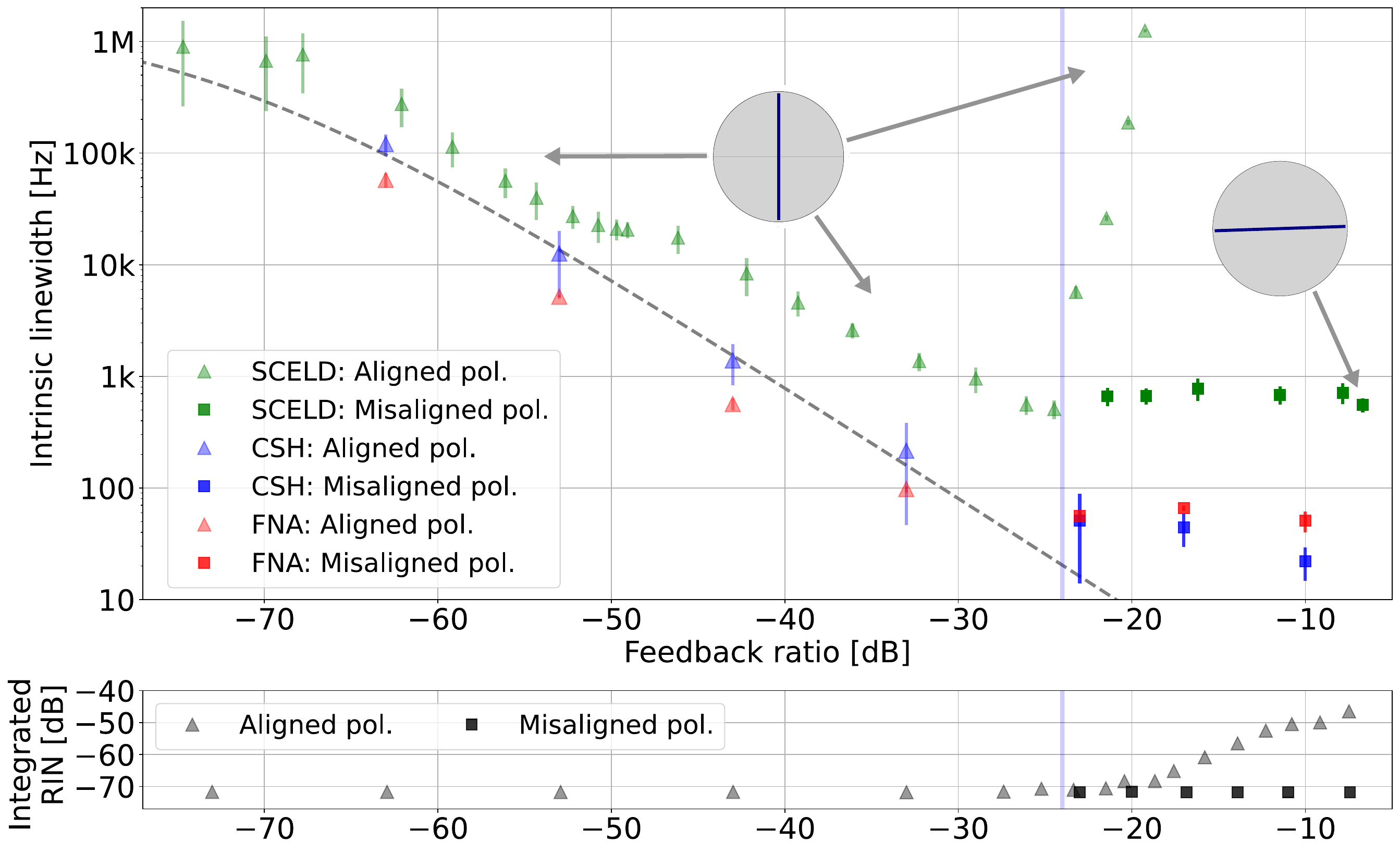}
        \label{fig: linewidth vs feedback}
    \end{subfigure}
    \hfill
    \begin{subfigure}[b]{0.46\textwidth}
        \vspace{-10pt}
        \caption{}
        \centering        
        \includegraphics[width=\textwidth]{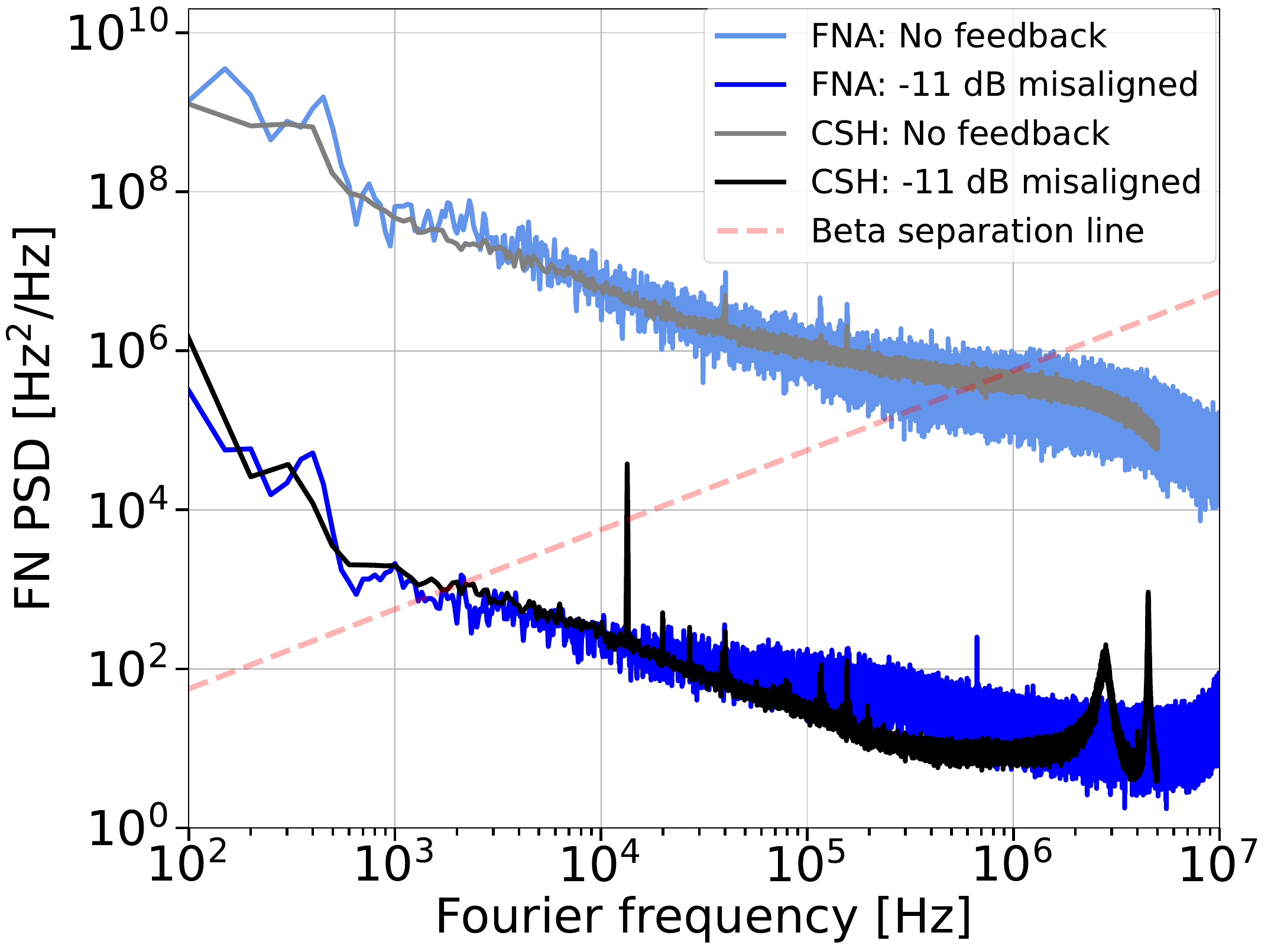}
    \label{fig:fn_psd}
    \end{subfigure}
    \hfill
    \begin{subfigure}[b]{0.50\textwidth}
        \vspace{-10pt}
        \caption{}
        \centering
        \includegraphics[width = \textwidth,height=0.68445\textwidth]{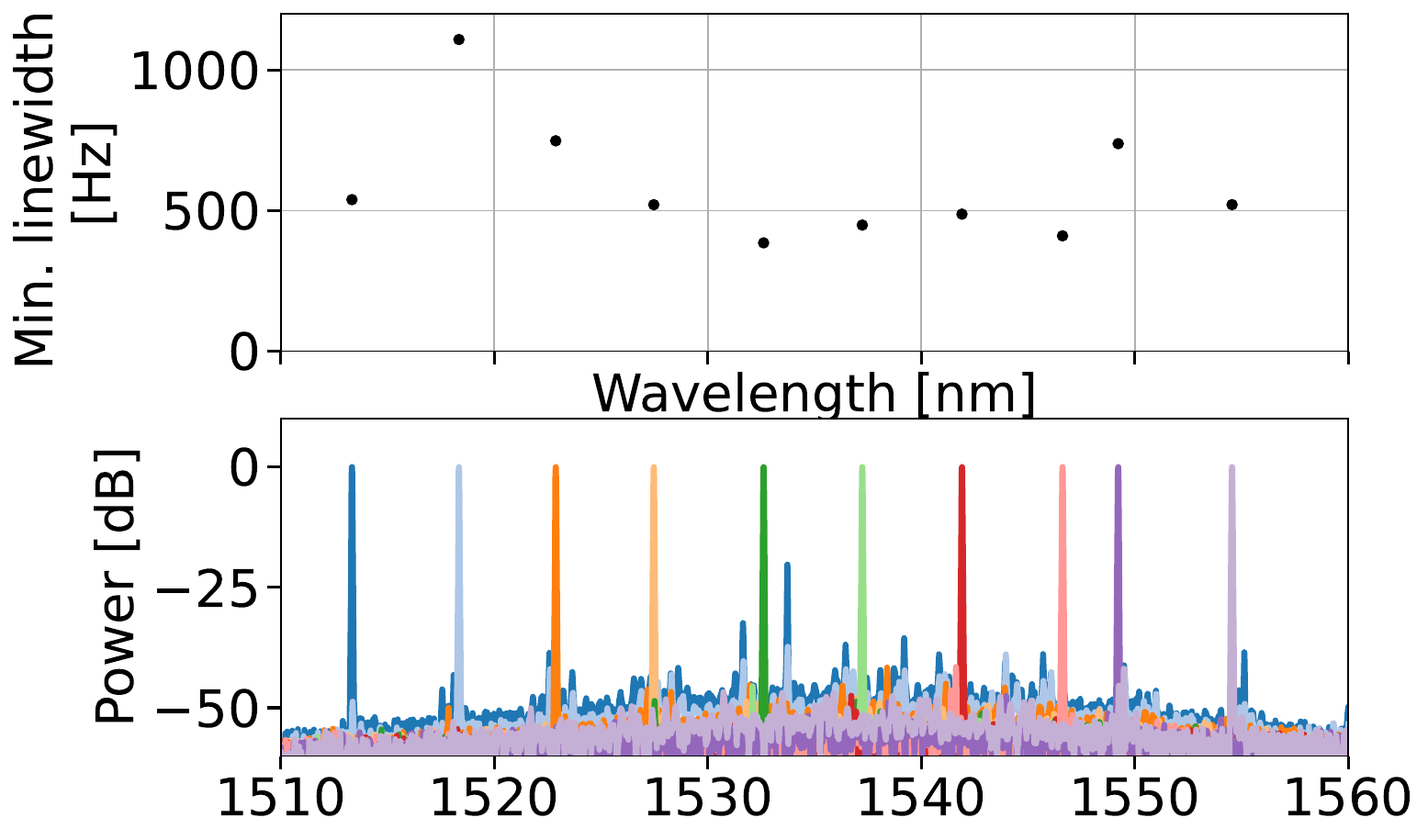}
        \label{fig: linewidth vs wavelength}
    \end{subfigure}
    \begin{subfigure}[t]{0.41\textwidth}
    \vspace{-30pt}
        \caption{}
        \centering
        \includegraphics[width = \textwidth,height=0.68445\textwidth]{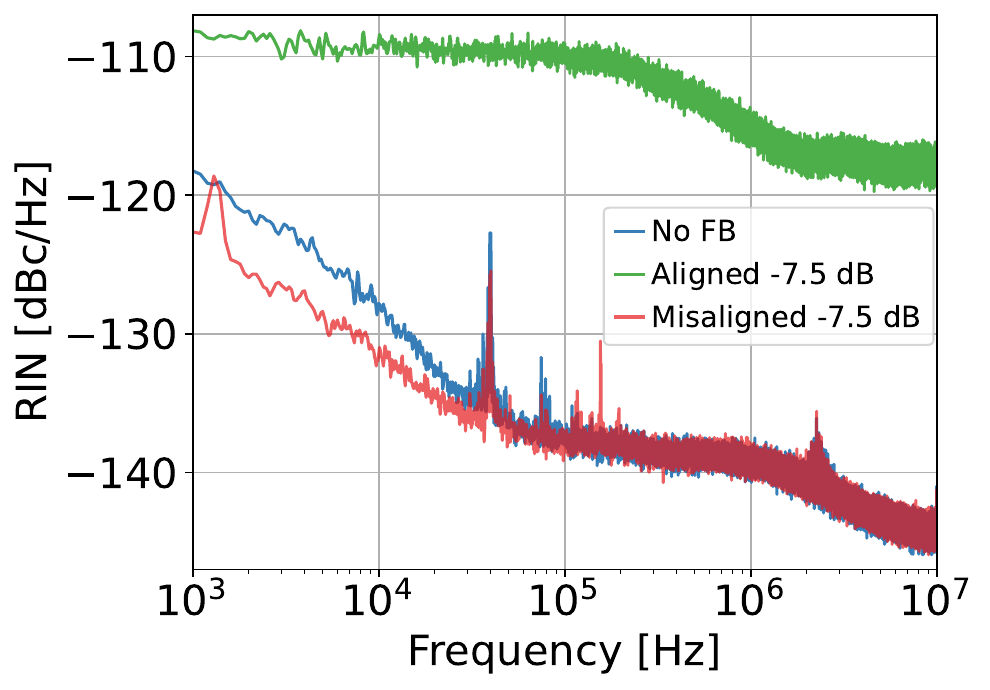}
        \label{fig: Highfinesse RIN PSD}
    \end{subfigure}
    \begin{subfigure}[t]{0.55\textwidth}
        \vspace{-30pt}
        \caption{}
        \centering        
        \includegraphics[width=\textwidth]{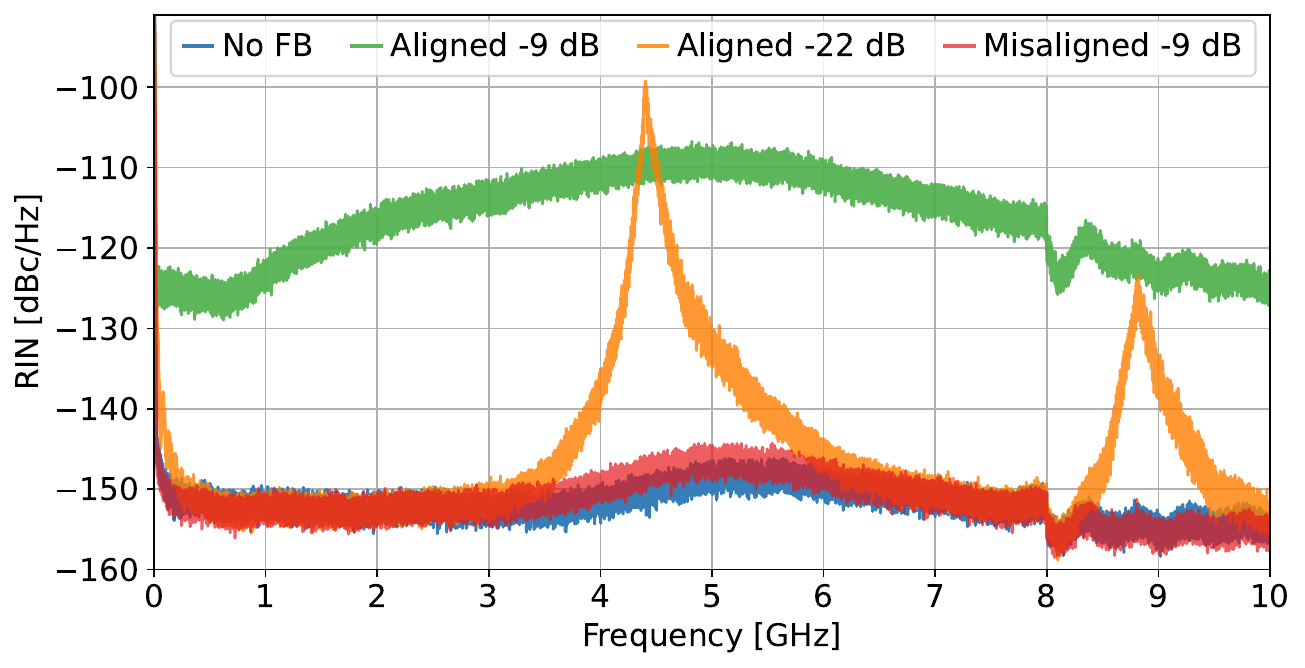}
    \label{fig:RIN 110mA gain no feedback}
    \end{subfigure}
    
    \vspace{-10pt}

    \caption{(a) Intrinsic linewidth (top) and integrated RIN (bottom) plotted against varying feedback ratios. The triangles indicate data points where the feedback polarization is aligned to the laser output, whereas the squares indicate misaligned polarizations. The insets show polarization ellipses at the polarimeter for aligned feedback polarization and one for strong ($-6$~dB) misaligned feedback. The purple line at $-24$~dB shows the onset of CC. (b) FN PSD for the laser without feedback and with -11 dB misaligned feedback using CSH (in grey/black) and the FNA (in blue). The beta separation line is plotted in red.
    (c) Top: Lowest measured SCELD linewidths of the different emission wavelengths. Bottom: OSA spectra for the different wavelengths.
    (d) Relative intensity noise (RIN) for the laser with strong ($-7.5$~dB) aligned feedback (green), strong ($-7.5$~dB) misaligned feedback (red) and no feedback (blue).
    (e) GHz range RIN spectra for various feedback ratios at 110mA gain current. No feedback (blue), medium ($-22$~dB) aligned feedback (orange), strong ($-9$~dB) aligned feedback (green), and strong ($-9$~dB), misaligned feedback (red). FB: feedback.}
    \label{fig:feedback}
\end{figure}

\subsection{Linewidth versus feedback power}

The primary objective of this work is 
to investigate the impact of polarization-controlled feedback on laser linewidth and stability, particularly under high feedback levels, pushing beyond the constraints of prior studies.
Using the VOA, we attenuated the feedback ratios by up to $70$~dB.
The laser operated with gain and SOA currents of $110$ mA and $40$ mA, respectively, producing 5 mW of fiber-coupled output power at a wavelength of 1549 nm. 
We obtained linewidth measurements using the SCELD method with various delays (as described in \cref{subsection: DSH measurement method}), ensuring single-mode operation with SMSRs exceeding $35$~dB (\cref{subsection: Stability}). For each feedback strength, we averaged 20 linewidth measurements, with uncertainties reflecting one standard deviation after removing outliers via the interquartile range method\cite{Sorensen_2024_Open-source}. FN spectra were likewise obtained using the CSH method and the FNA, with uncertainties representing one standard deviation based on 5 spectra. The resulting intrinsic linewidth measurements are shown against varying feedback power levels in \cref{fig: linewidth vs feedback} as either triangles (aligned feedback) or squares (misaligned feedback).

For the SCELD method (in green), the results reveal a linewidth reduction of more than three orders of magnitude, from MHz to sub-kHz, as feedback power increases.
For feedback ratios up to approximately $-24$~dB, increasing feedback levels resulted in progressively lower linewidths as indicated by the triangles.
At these feedback levels,  
the feedback polarization 
was aligned with the laser output,
to minimize the linewidth as predicted by \cref{eq:feedback}. However, maintaining this polarization alignment at higher feedback levels led to CC, where the linewidth significantly broadened. By misaligning the feedback polarization, a linewidth plateau of around $500$~Hz can be maintained for higher feedback levels, as represented by the squares in \cref{fig: linewidth vs feedback}.

Misaligning the polarization decreases the coupling strength $\beta$
between the feedback and output fields. Consequently, by reducing the polarization overlap between the two fields as $f_\mathrm{ext}$ increases, the effective feedback can be kept just below the onset of CC. At the highest feedback ratio of $-6$~dB, limited by the setup and losses, the feedback and laser output polarizations are nearly orthogonal, as evident from the insets in \cref{fig: linewidth vs feedback}. 
This allows us to obtain the minimum-achievable linewidth over a larger feedback range. Thus we bridge the need for fine-tuning the feedback level to being on the verge of CC, where small variations in the feedback intensity would be devastating to the linewidth.
Similar measurements for the FN spectrum (blue and red data points) reveal significantly lower intrinsic linewidths than the SCELD method. The linewidth plateaus for the CSH method and the FNA are $39\pm12$ Hz and $58\pm6$ Hz, respectively. The fact that the linewidth plateau for SCELD seems to be obtained already before the onset of CC, suggests that a measurement noise floor is reached. Further measurements using longer delays are needed to confirm that this is the case. Furthermore, the FNA measurements also seem to be noise floor limited at the plateau. This can be seen from measurements of FN spectra in \cref{fig:fn_psd}. The FNA specifies that linewidths measurements below $100$ Hz may be limited by the instrument noise floor. A theoretical fit to the pre-CC CSH data, based on \cref{eq:feedback} (grey dashed line), is plotted using the measured value of
$\Delta f_0 = 1.8 \pm 0.5$ MHz for the intrinsic linewidth without feedback and $\beta$ as the fitting parameter. The intersect of the theoretical line with the onset of CC predicts a linewidth plateau of $20$ Hz, slightly lower than the measured plateau. However, the linewidths at $-10$ dB feedback are within the uncertainty of the predicted plateau. We expect that this level can be achieved for the other CC mitigated data points with more precise tuning of the polarization.

The large uncertainties for data points below -60 dB in \cref{fig: linewidth vs feedback} suggests that unwanted feedback from Rayleigh backscattering or other reflections at this magnitude are present in the system. To investigate this, we measured the upper limit for unwanted feedback to be -60 dB. This was done by inserting a 1×2 splitter between the lensed fiber and the circulator (see \cref{fig: Meas setup 1}), directing one output to the lensed fiber and the other to a power meter. Port 3 of the circulator was unconnected. The difference when coupling port 2 of the circulator to the feedback loop was negligible, and cross-talk between the splitter’s outputs was at -60 dB, thus determining the upper limit.

Besides demonstrating the reduction in intrinsic linewidth, FN measurements were made to showcase that the method also reduces the effective linewidth. Fig. \ref{fig:fn_psd} shows laser FN spectra without feedback and with $-11$ dB misaligned feedback for the two different methods. The peak at around 10 kHz in the CSH is from the ESA background. Furthermore, the two peaks in the MHz range are due to external cavity modes from the different EOM sidebands. These peaks obstruct the noise floor that is hidden below. Consequently, the CSH noise floor is somewhat higher than it should be, which explains some of the discrepancy in the data points from Fig.~\ref{fig: linewidth vs feedback}. Using the beta separation line method \cite{DiDomenico:10}, the effective linewidth is reduced from $3.1$ MHz to $22$ kHz - a reduction by more than two orders of magnitude.

Next, by tuning the SG-DBR mirrors we modified the emission frequency across the full tunability range of the laser, as seen in the bottom of \cref{fig: linewidth vs wavelength}.
The fiber-coupled output power was increased to 10~mW by increasing SOA currents to 100 mA. For each wavelength, we measured the SCELD linewidth following the previously described method and optimized the polarization to yield the lowest possible linewidth. Results are shown in the top of \cref{fig: linewidth vs wavelength}, demonstrating that the laser maintains narrow linewidths across its tuning range (1513--1555 nm). Most wavelengths achieve a minimum linewidth of around 500 Hz for feedback ratios between $-15$~dB and $-20$~dB, which agrees with the linewidth plateau from \cref{fig: linewidth vs feedback}. We believe the discrepancy in linewidths across the different wavelengths can be reduced by more precise tuning of the mirror parameters and feedback polarization.
The mode lasing at 1513~nm has low power side modes between 1530~nm and 1540~nm indicating that 1513~nm is at the end of the tunability range.

These results show intrinsic linewidths two orders of magnitude lower than in similar studies on the U-shaped laser such as in Ref. \cite{Brusatori_Dual_wavelength_2023}, which examined the case of dual-wavelength emission. Compared to that study, we have incorporated polarization control, more accurate intrinsic linewidth estimation, and single-mode filtering based on SMSR. 
Our results are also at the forefront when comparing with recent findings for an SG-DBR laser that filtered the feedback using a Sagnac ring and two coupled ring resonators\cite{wang2024widely}. They achieved linewidth reductions by over three orders of magnitude (MHz to sub-kHz) with a comparable wavelength tuning range; however, our setup avoids the need for complex filtering components and reach sub-100~Hz intrinsic linewidths.

\subsection{RIN versus feedback power}

To further assess the laser's stability, we monitored RIN spectra up to 10 MHz, across all feedback ratios, using the ESA. Examples are shown on \cref{fig: Highfinesse RIN PSD} for strong feedback misaligned polarization (red), strong feedback aligned polarization (green), and no feedback (blue).
The measurements were taken with a 200~Hz resolution bandwidth and 50 averages. Our RIN measurements are limited by the noise floor from the photodetector. 
The RIN for the misaligned feedback case is shown to be lower than the no feedback case at the low frequencies until it reaches the detector noise floor.
The peaks seen between 30~kHz and 100~kHz most likely stem from technical noise in the laser. The spurs seen between 2~MHz and 3~MHz are due to TEC pick-up noise and are therefore not inherent to the laser.
Lastly, some feedback-induced spurs are seen between 100~kHz and 200~kHz for the misaligned polarization.

For each measurement, a single representative value was obtained by integrating over the spectra, from 1~kHz and onward, as reported on \cref{fig: linewidth vs feedback} (bottom graph).
These RIN measurements confirm that CC was successfully circumvented, as the expected sharp increase in RIN is mitigated\cite{Rasmussen_2019_Suppression}.

We performed further RIN measurements at the GHz range with a 10 MHz resolution bandwidth and 20 averages to confirm the mitigation of CC by monitoring the evolution of the relaxation oscillation frequency $f_\mathrm{RO}$ for different feedback levels and polarizations. The results are shown in \cref{fig:RIN 110mA gain no feedback}.
The notch at 8~GHz and subsequent ripples are due to background noise. Furthermore, the RIN measurements are detector-limited, due to the photodetector noise floor.
\begin{table}[tb]
    \centering
    \caption{Summary of laser properties depending on gain: resonance oscillation frequency $f_\mathrm{RO}$, the onset of CC for aligned feedback, and minimum linewidth obtained via SCELD method.}
    
    \begin{tabular}{c c c c}
        \hline
        Gain (mA) & $f_{\mathrm{RO}}$ (GHz) & Coherence collapse onset (dB) & Minimum linewidth (Hz)\\
        \hline
        $40$ & 1.3 & -50 & 20 k \\
        $60$ & 2.6 & -36 & 2.2 k  \\
        $80$ & 3.9 & -29 & 840 \\
        $110$ & 5.6 & -24 & 375\\
        
        \hline
    \end{tabular}
    \label{tab:coherence_vs_gain}
\end{table}

For no feedback (blue graph), $f_{\mathrm{RO}}$ is at 5.6 GHz, where it slightly decreases until the onset of CC. Increasing the feedback past CC while aligning the polarization to the laser output (orange graph) causes a further decrease in $f_\mathrm{RO}$ and an increase in the peak height and width. Furthermore, harmonics at integer multiples of $f_{\mathrm{RO}}$ appear, which are caused by the undamping of the relaxation oscillations, and are indicative of weak chaos \cite{porte2014similarity}.  A further increase in the aligned feedback (green graph) results in a broadening and increase of the RIN spectrum, which is indicative of strong chaos. For strong feedback, it is possible to mitigate the CC by misaligning the feedback polarization with the laser output polarization. This is evident from the strong, almost orthogonal feedback (red graph), which closely resembles the pre-CC spectrum (blue graph).

Strong chaos occurs when the feedback-induced lasing frequency shift overlaps with $f_\mathrm{RO}$\cite{porte2014similarity}. Thus, higher $f_\mathrm{RO}$ values indicate greater tolerance to coherent feedback, pushing the onset of CC. To demonstrate this, \Cref{tab:coherence_vs_gain} shows how the laser's properties depend on $f_\mathrm{RO}$, which increases with increasing gain currents. With increasing values of $f_\mathrm{RO}$, the onset of CC is pushed to higher feedback levels. The higher permissible coherent feedback and higher output powers allow us to achieve lower linewidths. In our case, going from $40$~mA to $110$~mA gain extends the onset of CC by $26$~dB and lowers the minimum achievable SCELD linewidth by nearly two orders of magnitude (from 20 kHz to 375 Hz).

\subsection{Outlook}
This work has shown an alternative method of achieving the minimum linewidth due to optical feedback. Tuning the polarization results in a minimum linewidth equivalent to tuning the feedback level to just before the onset of chaos. It remains to be seen whether misaligning the polarization will yield additional benefits to stability and linewidth. Further work is required to investigate this both theoretically and experimentally.
Additionally, the maximum feedback level achieved was $-6$~dB, i.e. 25~\%, due to the setup and loss therein. By using an optical amplifier we would be able to investigate the method at higher feedback powers. Here we would expect to see an upper limit for the effectiveness, as incoherent feedback eventually also leads to chaos \cite{ju2005dynamic}.

Increasing the length of the feedback loop (thus increasing $\tau_e$) would introduce narrower linewidths according to \cref{eq:feedback}, though it would also increase temperature fluctuations and decrease the mode spacing between the ECMs. Further work could also investigate this relationship between feedback loop length, linewidth, and laser stability. Future endeavors to increase the signal-to-noise ratio for RIN measurements will allow us to no longer be limited by the detector noise floor and allow comparison with state-of-the-art lasers.
Additionally, we would like to increase the laser's operational temperature using the Peltier element to measure the impact temperature has on stability. Furthermore, this shifts the tunability range to higher wavelengths and allows us to measure the performance across the C+L-band.

Lastly, future efforts to package the laser for improved thermal insulation, using polarization maintaining fiber in the feedback loop, and isolate the fiber feedback loop from ambient fluctuations are expected to significantly enhance the system's long-term stability. These measures will help mitigate temperature-induced drifts and environmental perturbations, ensuring more reliable performance over extended periods of time.

\section{Conclusion}
In this work, we have demonstrated a dual-cavity feedback mechanism that effectively reduces the linewidth to sub-100 Hz. In addition to this, we obtained sub-kHz linewidths across the wavelength tuning range of 42 nm using a U-shaped SG-DBR laser. We achieved this with a simple and cost-effective fiber-based external cavity and attained a reduction of more than three orders of magnitude compared to the solitary cavity linewidth with SMSRs exceeding $35$~dB. By carefully varying the fraction of coherent feedback using polarization control, the amount of feedback can be increased beyond the usual onset of CC while still maintaining narrow linewidths and stable RIN, extending the region of lowest achievable linewidth. We confirmed this result by analyzing the properties of the laser, such as the relaxation oscillation frequency, under injection of various ratios of feedback for different laser gains and polarizations evidencing the robustness of the method. These results open new avenues for developing widely tunable, narrow-linewidth lasers with strong optical feedback.

\vspace{1cm}

\begin{backmatter}
\bmsection{Funding} 
We acknowledge support from Innovation Fund Denmark (Grand Solutions projects GreenCOM and
CryptQ); Independent Research Fund Denmark (DFF, MOF project).

\bmsection{Acknowledgments}
The authors want to thank James Christensen at OE Solutions America, Inc. for his helpful
feedback and valuable discussions, which improved the quality of this work. We also appreciate OE Solutions for
supplying the U-shaped lasers and accompanying test equipment, as well as their support throughout our research.

\bmsection{Disclosures}
The authors declare no conflicts of interest.

\smallskip




\clearpage

\clearpage
\bmsection{Data Availability Statement}
Data underlying the results presented in this paper are available in Ref. \cite{Surrow_data_2025}.










\bigskip
 


\end{backmatter}
 



\bibliography{References}




\end{document}